\documentclass[11pt,preprint,showpacs]{revtex4}
\usepackage{graphicx}
\usepackage{dcolumn}
\usepackage{bm}
\usepackage{amsmath}
\usepackage{amssymb}
\usepackage{amsfonts}

\begin{document}

\title{On the discrepancies in the low energy neutron-deuteron breakup}

\author{H.~Wita{\l}a}
\affiliation{M. Smoluchowski Institute of Physics, Jagiellonian
University,
                    PL-30059 Krak\'ow, Poland}

\author{W.\ Gl\"ockle}
\affiliation{Institut f\"ur theoretische Physik II,
Ruhr-Universit\"at Bochum, D-44780 Bochum, Germany}

\date{\today}

\begin{abstract}
In view of recent  neutron-deuteron (nd) breakup  data for
 neutron-neutron (nn) and neutron-proton (np)
 quasi-free-scattering (QFS) arrangements and the large discrepancy found
between theoretical predictions and measured nn QFS cross sections,
 we analyze the sensitivity of the QFS cross sections to different partial
wave components of the nucleon-nucleon (NN) interaction.
 We found that the QFS cross section is strongly dominated by the $^1S_0$
and $^3S_1-^3D_1$ contributions. Because the standard three-nucleon force
(3NF) only weakly influence the QFS region, we conjecture, that it must be
the nn $^1S_0$ force component which is
 responsible for the discrepancy in the nn QFS peak. A stronger $^1S_0$ nn
force is required to bring theory and data into agreement.
 Such an increased strength of the nn interaction will, however, not help to
explain the nd breakup symmetric-space-star (SST) discrepancy. Further
experimental cross-checkings are required. \end{abstract}


\maketitle \setcounter{page}{1}

\section{Introduction}
 \label{intro}

The study of the nucleon induced deuteron breakup
 is a powerfull tool to test the nuclear Hamiltonian \cite{physrep96}. By
comparing theoretical predictions to breakup data in different
configurations of the outgoing nucleons not only the present day models of
two-nucleon (2N) interactions can be tested but also effects of
three-nucleon forces (3NF) can be  studied \cite{wit2001,kur2002}.

Despite spectacular successes in interpreting some breakup data based on a
3N Hamiltonian with free NN interactions supplemented by genuine 3NF's
\cite{physrep96,wit2001,kur2002}, some clear cut discrepancies between
theory and data exist and require further
 theoretical and experimental studies. Those discrepancies can be divided
into ones occuring at low and at higher energies. The last ones occur at
energies of the incoming nucleon above $E_N^{lab} \approx 100$~MeV and
 since the 3NF effects in the 3N continuum increase with energy
\cite{wit2001,kur2002}
 they can be probably traced  back to  neglected short-range terms of the
3NF.
 Also in some specific phase-space regions of the Nd breakup  effects of
relativity can play a role \cite{rel_wit1,rel_wit2,rel_skib1}

Low energy discrepancies were found in
  proton-deuteron (pd) pd and neutron-deuteron (nd)  breakup in
 some particular kinematical arrangements of the outgoing three nucleons
\cite{physrep96,tunl1,tunl2,strate1,siepe2,stephan1,gebhard1,church1,
church2,lubcke1,sagara,raup1,przyb1,ruan1}. The strongest discrepancies
between
theoretical cross sections
 and data occur in the nn QFS \cite{siepe2,ruan1,lubcke1} and in the
symmetrical nd space star (SST)
\cite{strate1,stephan1,tunl1,tunl2,gebhard1,church1,church2} geometries.

In the SST configuration three nucleons are flying away in the plane
perpendicular to the incoming beam in the c.m. system with momenta of equal
magnitude and a pairwise angular separation of  $120$ degrees.
 The low energy pd SST cross sections \cite{sagara,raup1,przyb1,zejma1}  are
slightly overestimated
 and the nd SST cross sections
\cite{tunl1,tunl2,strate1,stephan1,gebhard1,church1,church2} are clearly
underestimated by pure nd (the pp Coulomb force neglected) theoretical
predictions.
 The theoretical cross sections practically do not depend on the NN
potential used in the calculations~\cite{physrep96,kur2002}. They also do
not change significantly if any of the present day 3NF models is
included~\cite{physrep96,kur2002}.
 The observed diminishing of the disagreement between nd theory and pd SST
data with increasing proton energy: \cite{raup1} ($E_p^{lab}=13$~MeV),
\cite{przyb1} ($E_p^{lab}=19$~MeV), and \cite{zejma1} ($E_p^{lab}=65$~MeV),
leading to a quite good description of the  $65$~MeV
 pd data \cite{zejma1}, led to the conclusion, that probably the neglected
pp Coulomb force is responsible for the  low energy pd SST cross section
discrepancies. However, recent breakup calculations with the long-range
Coulomb
force included revealed only a  small lowering of the SST cross section by
the pp Coulomb force \cite{deltuva2}. Much larger discrepancies up to
$\approx 30 \%$ have been found for the  nd SST. The data come from
independent measurements, performed at the same energy and with different
experimental arrangements: $E_n^{lab}=10.3$~MeV
\cite{stephan1,gebhard1,church1}, $E_n^{lab}=13$~MeV
\cite{strate1,tunl1,tunl2}, $E_n^{lab}=16$~MeV \cite{church2}.

The QFS refers to a situation where one of the nucleons is at rest in the
laboratory system. In the pd breakup the np or pp QFS configurations are
possible, while for the nd breakup np or nn can form a quasi-freely
scattered pair. The cross sections for pp QFS in pd breakup: \cite{raup1}
($E_p^{lab}=13$~MeV),
 \cite{przyb1} ($E_p^{lab}=19$~MeV), and \cite{allet1}
($E_p^{lab}=65$~MeV), and for nn QFS: \cite{siepe2} ($E_n^{lab}=26$~MeV),
\cite{ruan1} ($E_n^{lab}=25$~MeV), and
 \cite{lubcke1} ($E_n^{lab}=10.31$~MeV), and np QFS \cite{siepe2} in nd
breakup
 have been measured and the picture resembles that for SST:
 also here the pp QFS cross sections are overestimated by the nd theory
while the nn QFS cross sections are clearly underestimated by $\approx 20
\%$.
  Suprisingly, when instead of the nn pair the np pair is quasi-freely
scattered, the theory follows
 nicely the np QFS cross section data  taken in the nd breakup measurement
\cite{siepe2}.
 Also the nd theory provides  QFS cross sections which are highly
independent from the realistic NN potential used in the calculations and
they practically do not change when any of the present day 3NF's is
included~\cite{physrep96,kur2002}. Again, the discrepancies between nd
theory and pp QFS cross sections diminish
  with increasing energy \cite{raup1,przyb1,allet1},
 indicating  that the neglected pp Coulomb  force is probably responsible
for them.
 Also here the pd breakup calculations show that the pp Coulomb force only
slightly lowers the nd theory in the pp QFS region \cite{deltuva2}.

The good description of the np QFS cross section data from the
$E_n^{lab}=26$~MeV nd breakup measurement \cite{siepe2} contrasts with the
drastic discrepancy to  the nn QFS cross section data taken in the same
experiment \cite{siepe2}. This
 prompted us to look closely how different NN force components contribute to
QFS cross sections.
This is outlined in section II, while section III deals with the SST cross
sections. We summarize in
section IV.

  \section{Sensitivity test for QFS cross sections}
\begin{figure}
\includegraphics[scale=0.8]{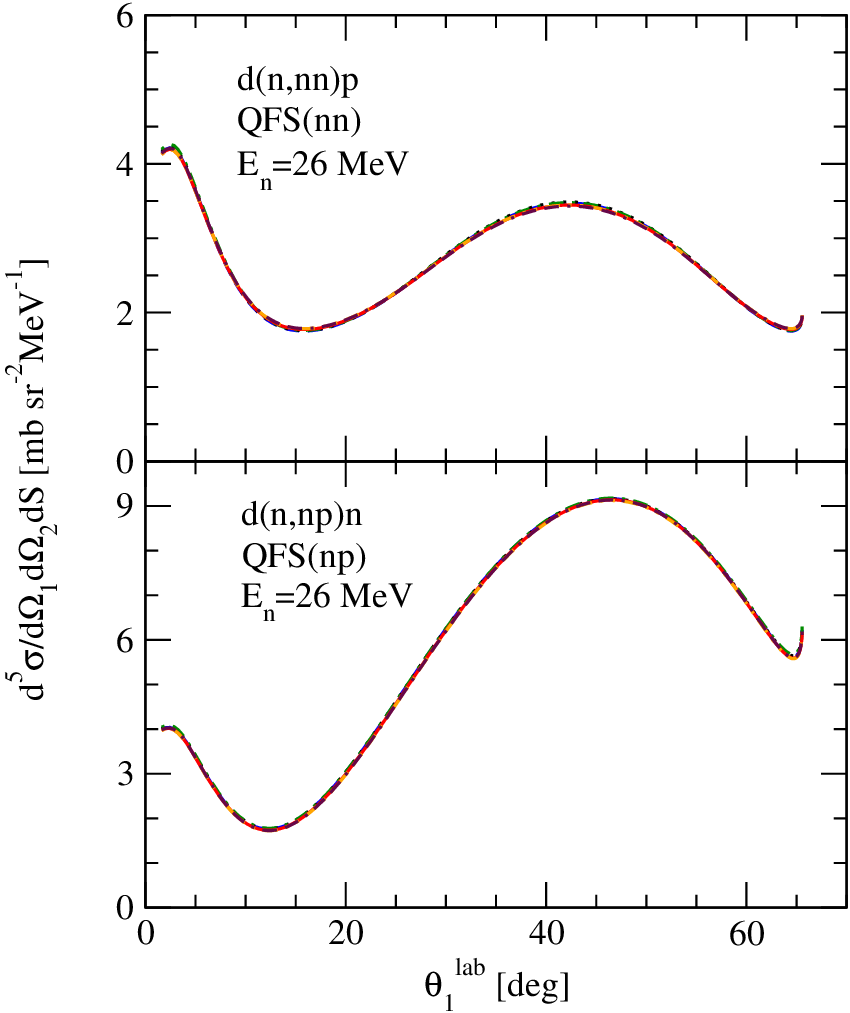}
\caption{(color online) The cross section $d^5\sigma/d\Omega_1d\Omega_2dS$
for the $E_n^{lab}=26$~MeV nd breakup reaction $d(n,nn)p$ (upper panel) and
$d(n,np)n$ (lower panel) under exact QFS condition (the momentum of the
undetected third nucleon $\vec p_3~^{lab} = 0$: nn QFS for the upper and np
QFS for the lower panel) as a function of the laboratory angle of the
 detected neutron. The (overlapping)  lines
correspond to different underlying dynamics: CD Bonn - dashed (blue), Nijm I
- dotted (black), Nijm II - dashed-dotted (green), CD Bonn+TM99 - solid
(red), Nijm I +TM99 - dashed-double-dotted (orange), Nijm II + TM99 -
double-dashed-dotted (maroon). All partial waves with 2N total angular
momenta up to $j_{max}=5$ have been included.
}
\label{fig1}
\end{figure}

\begin{figure}
\includegraphics[scale=0.8]{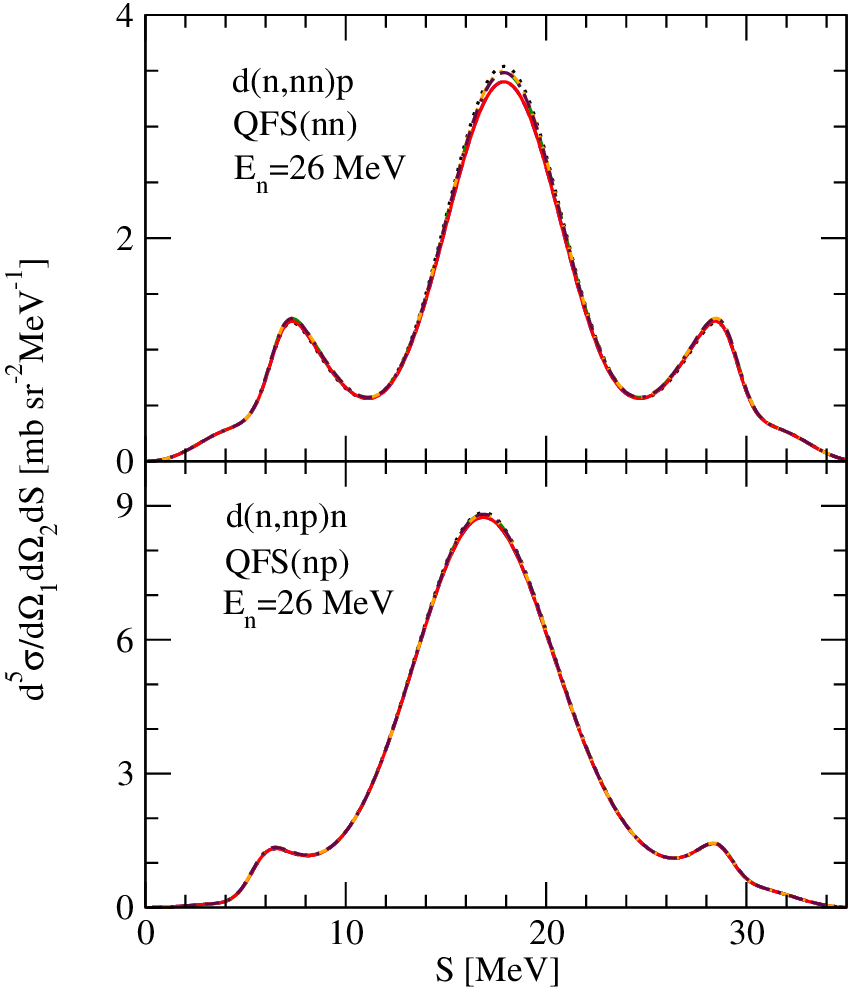}
\caption{(color online) The cross section $d^5\sigma/d\Omega_1d\Omega_2dS$
for the $E_n^{lab}=26$~MeV nd breakup reaction $d(n,nn)p$ (upper panel) and
$d(n,np)n$ (lower panel) as a function of the S-curve length for two
complete configurations of Ref.~\cite{siepe2}.
QFS nn refers to the angles of two neutrons: $\theta_1=\theta_2=42^o$
 and QFS np refers to the angle $\theta_1=39^o$   of  the detected neutron
 and  $\theta_2=42^o$ for the proton. In both cases $\phi_{12}=180^o$. For
the description of lines see Fig.~\ref{fig1}.
}
\label{fig2}
\end{figure}

The tests are performed using the  (semi)phenomenological NN potentials
alone (AV18 \cite{av18},
CD Bonn \cite{cdbonn}, Nijm1 and Nijm2 \cite{nijm})  or combining them
 with the $2\pi$-exchange TM99  3NF \cite{TM,tm99}.
 For details of our approach to solve the 3N Faddeev equation, which is
based
 on momentum space partial wave decomposition,
 we refer to \cite{physrep96,gloeckle83}.
As an example we checked that at the  energy $E_n^ {lab} = 26$~ MeV  the np
and nn QFS cross section are practically insensitive to the exchange between
different realistic NN potentials and that the effects of 3NF's are
negligible
 (see Fig.~\ref{fig1} and \ref{fig2}).

\begin{figure}
\includegraphics[scale=0.8]{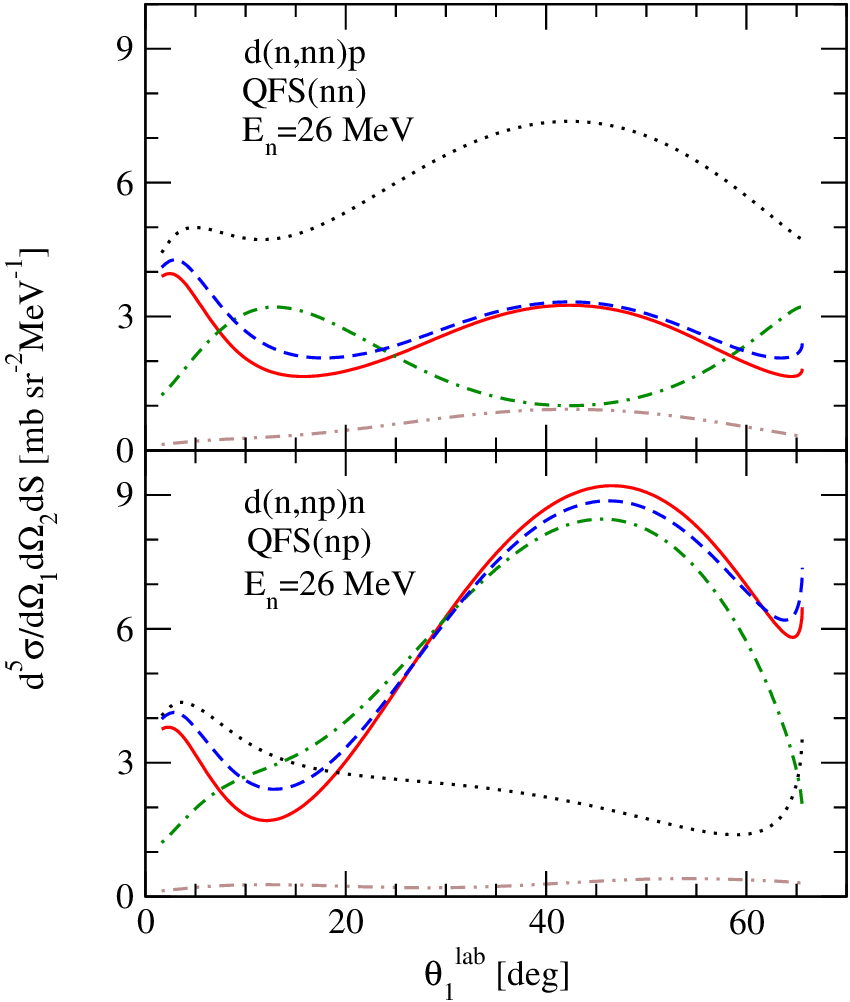}
\caption{(color online) The cross section $d^5\sigma/d\Omega_1d\Omega_2dS$
for the $E_n^{lab}=26$~MeV nd breakup reaction $d(n,nn)p$ (upper panel) and
$d(n,np)n$ (lower panel) under exact QFS condition (the momentum of the
undetected third nucleon $\vec p_3~^{lab} = 0$: nn QFS for the upper  and np
QFS for the lower panel) as a function of the laboratory angle of
 the  detected neutron. The different lines show contributions from
different NN force components. The solid (red) line is the full result based
on the CD Bonn potential \cite{cdbonn} and all partial waves with 2N total
angular momenta up to $j_{max}=5$ included. The dotted (black),
dashed-dotted (green), and dashed (blue) lines result when only
contributions from $^1S_0$, $^3S_1-^3D_1$, and $^1S_0+^3S_1-^3D_1$ are kept
when calculating the cross sections. The dashed-double-dotted (brown) line
presents the contribution of all partial waves with the exception of $^1S_0$
and $^3S_1-^3D_1$.
}
\label{fig3}
\end{figure}

\begin{figure}
\includegraphics[scale=0.8]{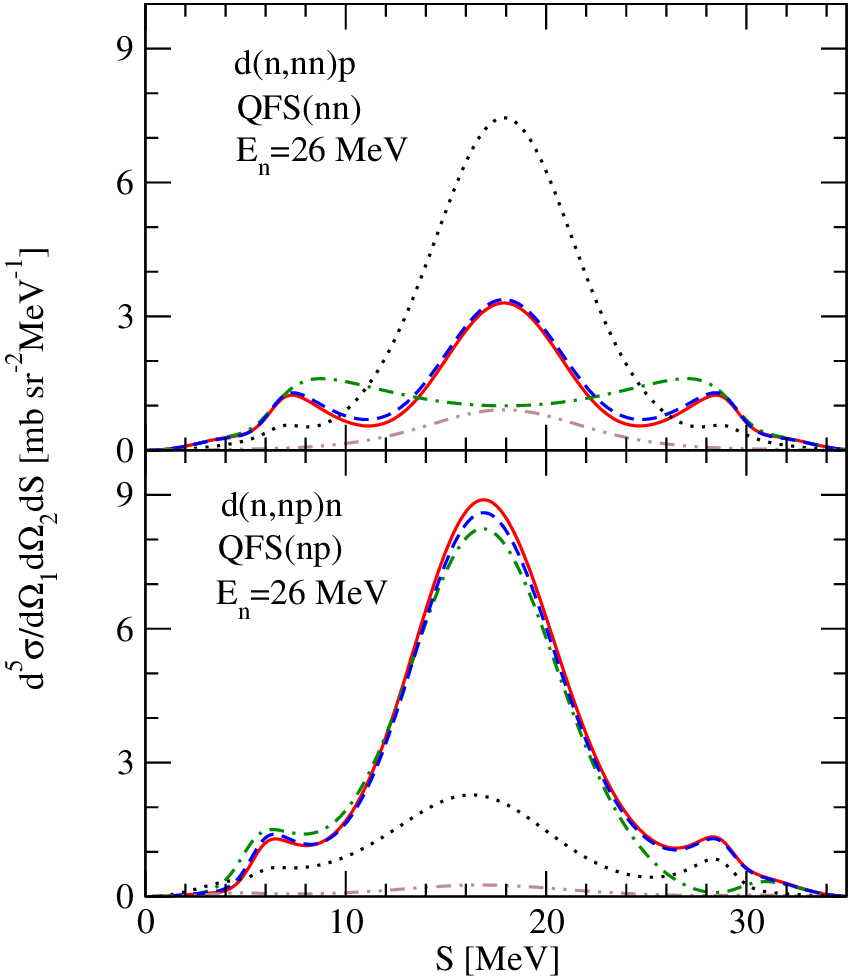}
\caption{(color online) The cross section $d^5\sigma/d\Omega_1d\Omega_2dS$
for the $E_n^{lab}=26$~MeV nd breakup reaction $d(n,nn)p$ (upper panel) and
$d(n,np)n$ (lower panel) as a function of the S-curve length for two
complete configurations of Ref.~\cite{siepe2}.
QFS nn refers to the angles of two neutrons: $\theta_1=\theta_2=42^o$
 and QFS np refers to the angle $\theta_1=39^o$ of the detected   neutron
 and  $\theta_2=42^o$ for the proton. In both cases $\phi_{12}=180^o$. For
description of lines see Fig.~\ref{fig3}.
}
\label{fig4}
\end{figure}

In Fig.~\ref{fig1}  we show the cross section
$d^5\sigma/d\Omega_1d\Omega_2dS$ for the nd breakup reaction $d(n,N_1
N_2)N_3$ under exact QFS condition ($\vec p_3~^{lab} = 0$) as a function of
the laboratory angle $\theta_1$ of the outgoing nucleon $N_1$. In
Fig.~\ref{fig2}  we show nn and np QFS cross sections for configurations
measured in Ref. \cite{siepe2}.
 Since the QFS cross section is practically insensitive to the action of a
3NF the reason for the nearly $20 \%$ underestimation of the nn QFS cross
section data found in \cite{siepe2}  must be traced to the underlying NN
force. At low energy the largest contribution should be provided by the
S-wave components
 of the NN potential. In case of  free np and nn scattering these are the
$^1S_0$(np)+$^3S_1-^3D_1$ and $^1S_0$(nn) contributions, respectively. For
the nd breakup and  np or nn QFS configurations that dominance can also be
expected. However, due to the presence of an additional nucleon in the
incoming deuteron, for nn QFS also a smaller contribution from the
$^3S_1-^3D_1$ np interaction  is
 possible. In Figs.~\ref{fig3} and \ref{fig4} we show contributions
 to the np and nn QFS cross sections from $^1S_0$, $^3S_1-^3D_1$, and from
all other components of the NN interaction up to a two-body total angular
momentum $j=5$.
 It is clearly seen, that at low energies the QFS peak is build
predominantely
 from $^1S_0$ and $^3S_1$ components, with practically negligible
contributions from higher angular momentum states. It is also seen that
quite different pictures arise for np and nn QFS. While for np QFS
$^3S_1-^3D_1$ is the most dominant  contribution, for nn QFS
 it is  the $^1S_0$ force which  contributes decisively.
 If the disagreement between data and theory in the nn QFS peak has its
source in one or both of these  two force components  it must be the $^1S_0$
nn force. The $^1S_0$ nn force is  determined up to now only indirectly due
to lack of free nn data. Both $^1S_0$  and $^3S_1-^3D_1$ np forces are well
determined by np scattering data and by the deuteron properties. Also the
agreement for np QFS found in \cite{siepe2} further  supports the
possibility of a wrong nn $^1S_0$ interaction. Namely if QFS resembles
 free nucleon-nucleon scattering then the $^1S_0$ contribution to the nn QFS
should be mostly of the $^1S_0$ nn type whereas the $^1S_0$ contribution to
the np QFS is mostly given by the $^1S_0$ np force component. That would
imply
 that the nn QFS
 is a powerfull tool to study $^1S_0$ nn force component. 
Our numerical studies revealed, that at low energies higher order
rescatterings are important.
 As a result quasi-free-scattering
 does not resemble exactly free scattering and in the nn QFS peak also
contributions
 from  $^3S_1-^3D_1$ and
 $^1S_0$ np are present as well as in the np QFS peak there is also a small
admixture
 from  $^1S_0$ nn together with admixture in both, nn and np QFS's,
 from higher partial wave components.
 Nevertheless the low energy  dominance of the $^1S_0$ nn contribution in
the nn QFS peak and
 the dominance of the $^1S_0$ and $^3S_1-^3D_1$ np forces in the np QFS peak
remains true also in the presence of rescatterings.

\begin{figure}
\includegraphics[scale=0.9]{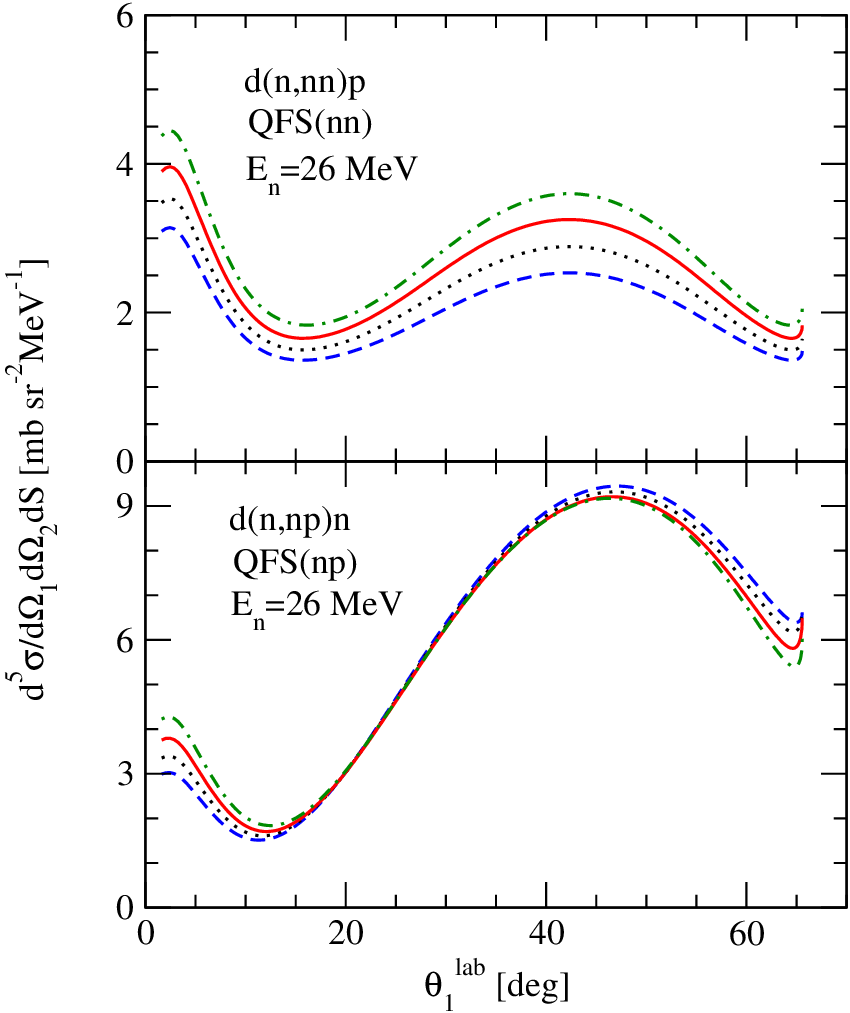}
\caption{(color online) The cross section $d^5\sigma/d\Omega_1d\Omega_2dS$
for the $E_n^{lab}=26$~MeV nd breakup reaction $d(n,nn)p$ (upper panel) and
$d(n,np)n$ (lower panel) under exact QFS condition (the momentum of the
undetected third nucleon $\vec p_3~^{lab} = 0$: nn QFS for the upper and np
QFS for the lower panel) as a function of the laboratory angle of the
 detected neutron. The lines show the sensitivity of the QFS cross
section to  changes of the nn $^1S_0$ force component. Those changes were
induced by multiplying the $^1S_0$ nn matrix element of the CD Bonn
potential by a factor $\lambda$. The solid (red) line is the full result
based on the original CD Bonn potential \cite{cdbonn} and all partial waves
with 2N total angular momenta up to $j_{max}=5$ included. The dashed (blue),
dotted (black), and dashed-dotted (green) lines correspond to $\lambda=0.9$,
$0.95$, and $1.05$, respectively. }
\label{fig5}
\end{figure}

\begin{figure}
\includegraphics[scale=0.9]{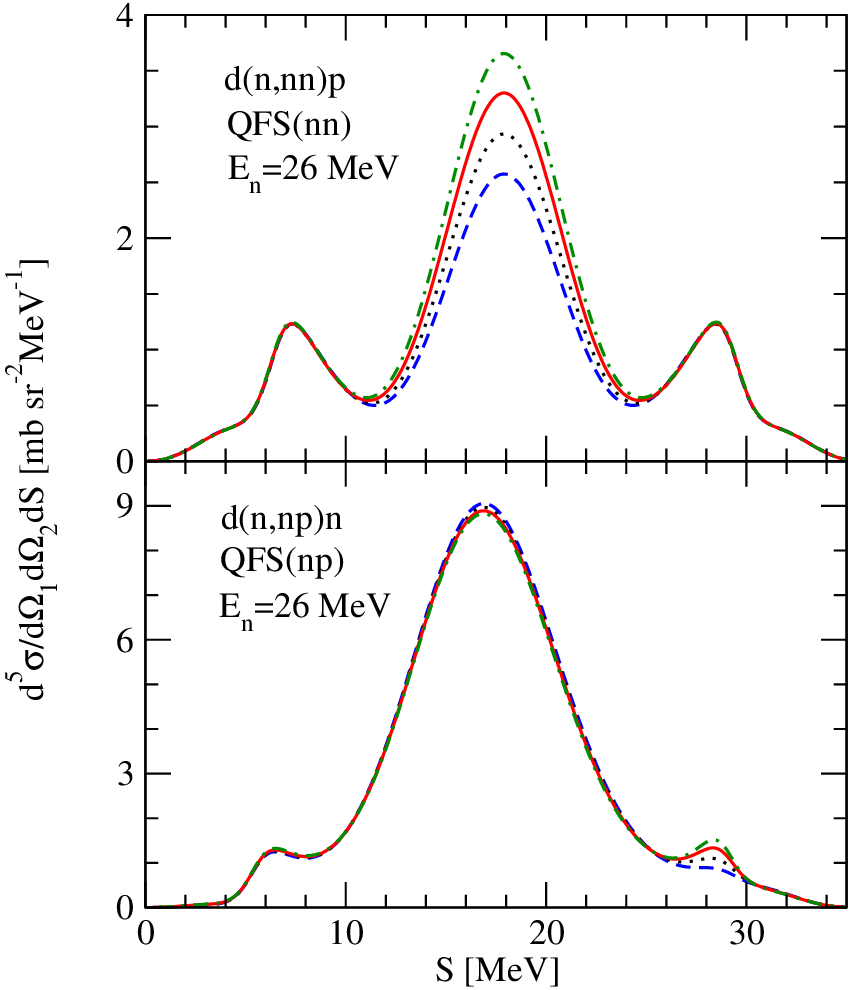}
\caption{(color online) The cross section $d^5\sigma/d\Omega_1d\Omega_2dS$
for the $E_n^{lab}=26$~MeV nd breakup reaction $d(n,nn)p$ (upper panel) and
$d(n,np)n$ (lower panel) as a function of the S-curve length for two
complete configurations of Ref.~\cite{siepe2}.
QFS nn refers to the angles of two neutrons: $\theta_1=\theta_2=42^o$
 and QFS np refers to the angle of detected  neutron $\theta_1=39^o$ and
proton  $\theta_2=42^o$. In both cases $\phi_{12}=180^o$. For description of
lines see Fig.~\ref{fig5}.
}
\label{fig6}
\end{figure}

That indeed the nn QFS is extremely sensitive to the $^1S_0$ nn force
component  we demonstrate in Figs.~\ref{fig5} and \ref{fig6} by showing
 the changes of the QFS cross section induced by changes of the nn $^1S_0$
interaction. In this  study we chosse a very simple device for that
change, namely multiplying the $^1S_0$ nn matrix element of the CD Bonn
potential by a factor $\lambda$. The result is, that  the nn QFS undergoes
 significant variations while the
 np QFS cross sections are practically unchanged.

\begin{figure}
\includegraphics[scale=0.8]{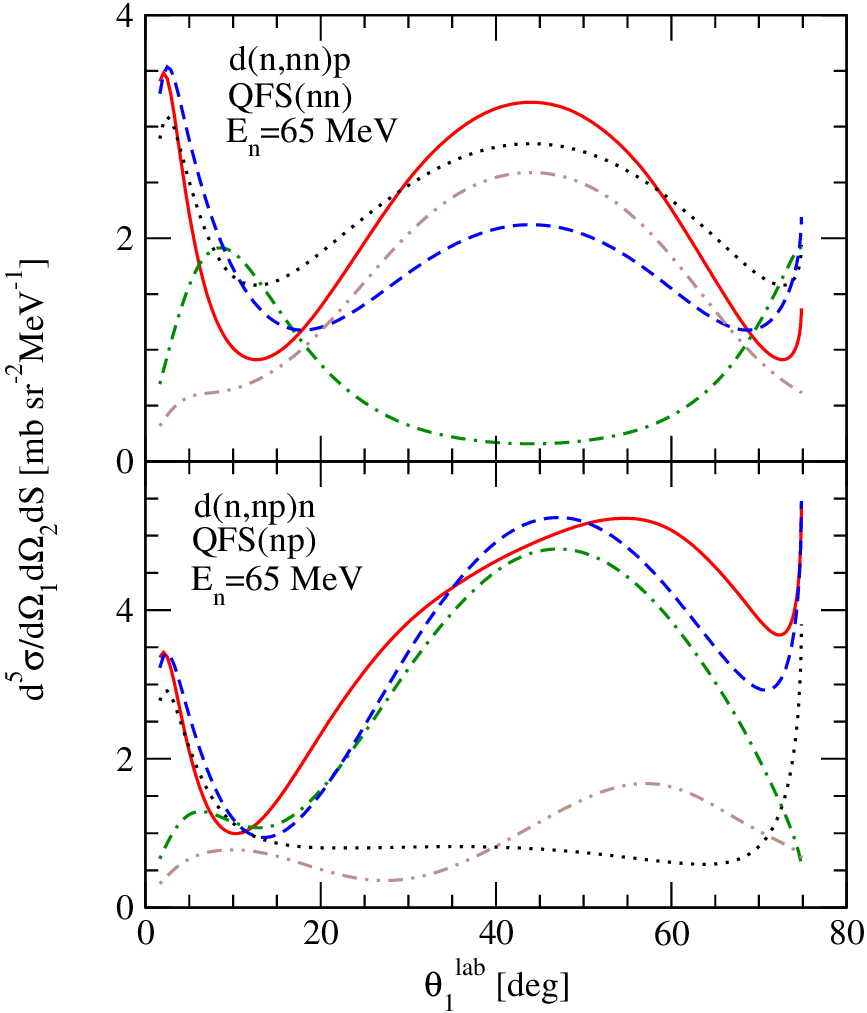}
\caption{(color online) The cross section $d^5\sigma/d\Omega_1d\Omega_2dS$
for the $E_n^{lab}=65$~MeV nd breakup reaction $d(n,nn)p$ (upper panel) and
$d(n,np)n$ (lower panel) under exact QFS condition (the momentum of the
undetected third nucleon $\vec p_3~^{lab} = 0$: nn QFS for the upper  and np
QFS for the lower panel) as a function of the laboratory angle of the
detected neutron. The lines show contributions from different NN force
components like in  Fig.\ref{fig3}.
}
\label{fig7}
\end{figure}

\begin{figure}
\includegraphics[scale=0.9]{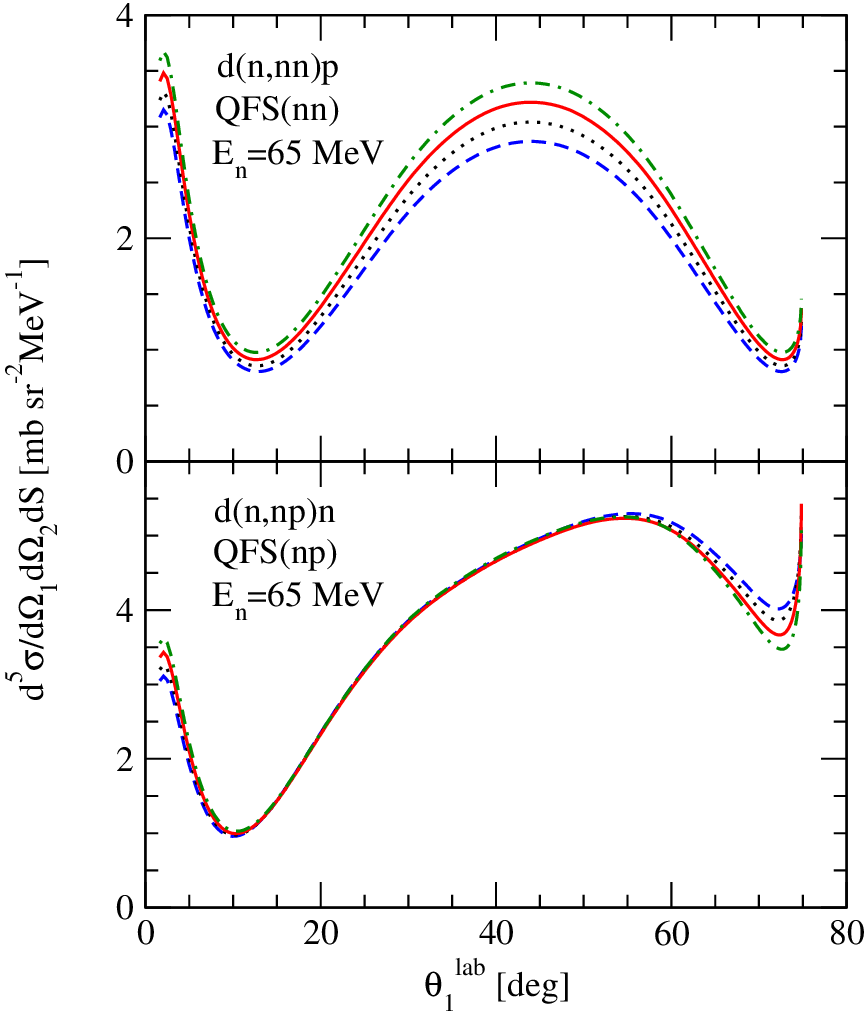}
\caption{(color online) The cross section $d^5\sigma/d\Omega_1d\Omega_2dS$
for the $E_n^{lab}=65$~MeV nd breakup reaction $d(n,nn)p$ (upper panel) and
$d(n,np)n$ (lower panel) under exact QFS condition (the momentum of the
undetected third nucleon $\vec p_3~^{lab} = 0$: nn QFS for the upper and np
QFS for the lower panel) as a function of the laboratory angle of the
 detected neutron. The lines show sensitivity of the QFS cross section to
the changes of the nn $^1S_0$ force component. Those changes were induced by
multiplying the $^1S_0$ nn matrix element of the CD Bonn potential by a
factor $\lambda$ like in  Fig.\ref{fig5}.
}
\label{fig8}
\end{figure}

Going to higher energies the QFS cross sections become sensitive also to
higher angular momentum  components of the NN interaction and the dominance
of S-waves is lost. Also the sensitivity of the nn QFS peak to changes of
the $^1S_0$ nn interaction is significantly reduced. This is demonstrated
 in Figs.~\ref{fig7} and  \ref{fig8}  at $E_n^{lab}=65$~MeV.

One can ask which change of the $^1S_0$ nn force is required to bring the nn
QFS theoretical cross sections into agreement with the data. To remove the
$18 \%$ discrepancy found in \cite{siepe2} for  nn QFS would require a value
of
 $\lambda \approx 1.08$. Such an increased strength of the $^1S_0$ nn
interaction would drastically decrease $^1S_0$ nn scattering length $a_{nn}$
to $a_{nn}^{^1S_0}(\lambda =1.08)=-135$~fm and would lead to a nearly  bound
state of two neutrons.

\section{ Sensitivity test for SST cross sections}

\begin{figure}
\includegraphics[scale=0.85]{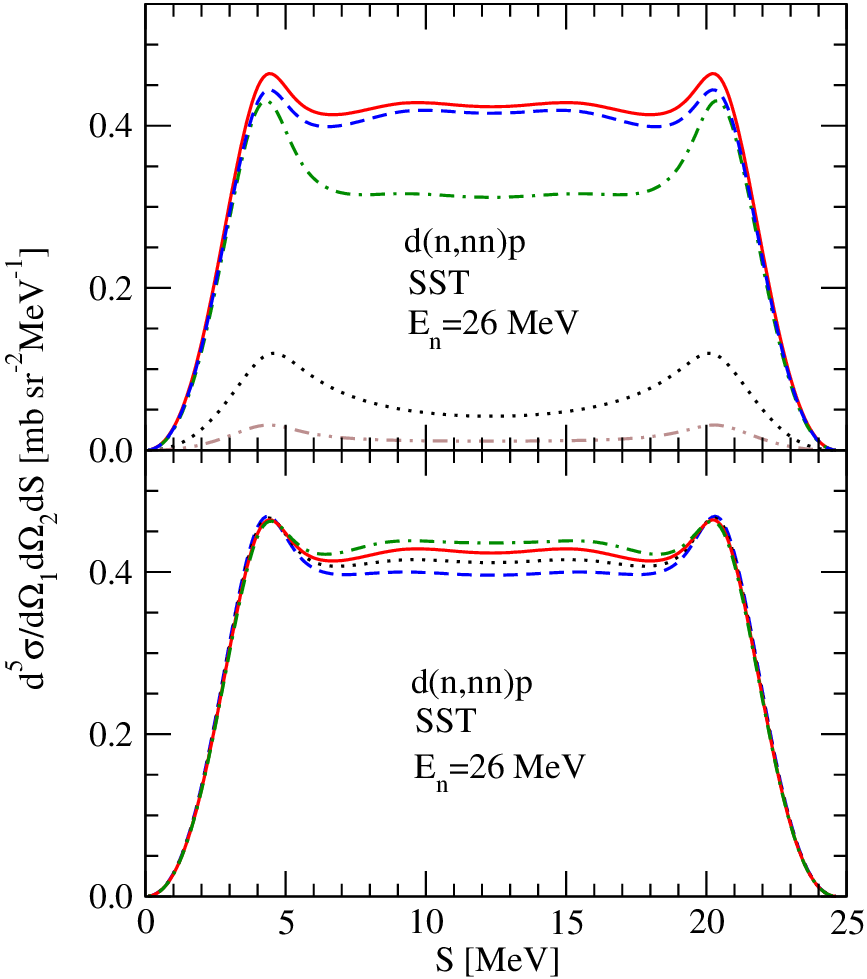}
\caption{(color online) The cross section $d^5\sigma/d\Omega_1d\Omega_2dS$
for the $E_n^{lab}=26$~MeV nd breakup reaction $d(n,nn)p$  as a function of
the S-curve length for SST  configuration with the lab. angles of two
detected neutrons $\theta_1=\theta_2=52.8^o$ and $\phi_{12}=180^o$. In the
upper panel the lines show contributions from different NN force components.
The solid (red) line is the full result based on the CD Bonn potential
\cite{cdbonn} and all partial waves with 2N total angular momenta up to
$j_{max}=5$ included. The dotted (black), dashed-dotted (green), and dashed
(blue) lines result when only contributions from $^1S_0$, $^3S_1-^3D_1$, and
$^1S_0+^3S_1-^3D_1$ are kept when calculating the cross sections. The
dashed-double-dotted (brown) line presents contribution of all partial waves
with the exception of $^1S_0$ and $^3S_1-^3D_1$. In the lower panel the
lines show sensitivity of the SST cross section to the changes of the nn
$^1S_0$ force component. Those changes were induced by multiplying the
$^1S_0$ nn matrix element of the CD Bonn potential by a factor $\lambda$.
The solid (red) line is the full result based on the original CD Bonn
potential \cite{cdbonn} and all partial waves with 2N total angular momenta
up to $j_{max}=5$ included. The dashed (blue), dotted (black), and
dashed-dotted (green) lines correspond to $\lambda=0.9$, $0.95$, and $1.05$,
respectively. }
\label{fig9}
\end{figure}

\begin{figure}
\includegraphics[scale=0.84]{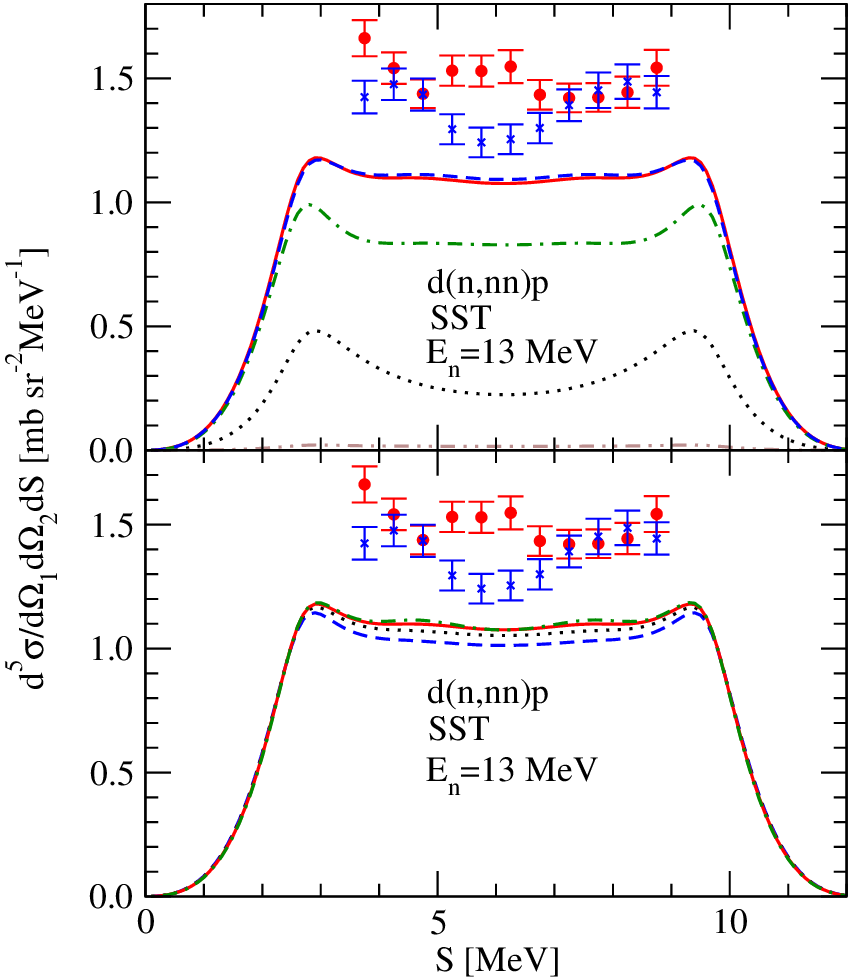}
\caption{(color online) The cross section $d^5\sigma/d\Omega_1d\Omega_2dS$
for the $E_n^{lab}=13$~MeV nd breakup reaction $d(n,nn)p$  as a function of
the S-curve length for SST  configuration with the lab. angles of two
detected neutrons $\theta_1=\theta_2=50.5^o$ and $\phi_{12}=180^o$. In the
upper  panel the lines show contributions from different NN force
components. The solid (red) line is the full result based on the CD Bonn
potential \cite{cdbonn} and all partial waves with 2N total angular momenta
up to $j_{max}=5$ included. The dotted (black),
 dashed-dotted (green), and dashed (blue) lines result when only
contributions from $^1S_0$, $^3S_1-^3D_1$, and $^1S_0+^3S_1-^3D_1$ are kept
when calculating the cross sections. The dashed-double-dotted (brown) line
presents contribution of all partial waves with exception of $^1S_0$ and
$^3S_1-^3D_1$. In the lower panel the lines show sensitivity of the SST
cross section to the changes of the nn $^1S_0$ force component. Those
changes were induced by multiplying the $^1S_0$ nn matrix element of the CD
Bonn potential by a factor $\lambda$. The solid (red) line is the full
result based on the original CD Bonn potential \cite{cdbonn} and all partial
waves with 2N total angular momenta up to $j_{max}=5$ included. The dashed
(blue), dotted (black), and dashed-dotted (green) lines correspond to
$\lambda=0.9$, $0.95$, and $1.05$, respectively. The solid dots and x-ses
are nd data of Ref. \cite{tunl1,tunl2} and \cite{strate1}, respectively. }
\label{fig10}
\end{figure}

Here we perform corresponding studies for the nd SST break up configuration.
Even such a drastic  increase
 of the   nn $^1S_0$ force strength by a factor $\lambda \approx 1.08$
 would  not remove the discrepancy for the nd SST configuration.
 Also here the main contributions to the cross section come from the $^1S_0$
and $^3S_1-^3D_1$ components. Contrary to nn QFS here the $^3S_1-^3D_1$
 is the dominant component.
 The building up of the SST cross sections  in $E_n^{lab}=26$~MeV and
$E_n^{lab}=13$~MeV nd breakup
 by components of the NN force is shown in the upper panels of
Figs.~\ref{fig9} and  \ref{fig10}, respectively,  and the  sensitivity to
the nn $^1S_0$ force in the lower
 panels of Figs.~\ref{fig9} and \ref{fig10}. It is seen that changes which
would provide an explanation for the nn QFS have practically no effect on
the SST discrepancy.

\section{Summary}
Summarizing, the QFS Nd breakup  cross sections are influenced only slightly
by present day 3NF's.
 Thus they are not responsible for the large differences between theoretical
predictions and data found for the nd breakup  nn QFS cross sections. The
study of the sensitivity of these cross sections to different NN force
components has shown, that in order to explain these discrepancies a
stronger nn $^1S_0$ force is required, leading however in our simplistic
modification to an unrealistic nn scattering length.
 The SST discrepancy is even more serious.
 Further measurements are required to solidify the experimental situation.

\section*{Acknowledgments}
This work was supported by the Polish 2008-2011 science funds as the
 research project No. N N202 077435.
It was also partially supported by the Helmholtz
Association through funds provided to the virtual institute ``Spin
and strong QCD''(VH-VI-231)  and by
  the European Community-Research Infrastructure
Integrating Activity
``Study of Strongly Interacting Matter'' (acronym HadronPhysics2,
Grant Agreement n. 227431)
under the Seventh Framework Programme of EU.
 The numerical
calculations have been performed on the
 supercomputer cluster of the JSC, J\"ulich, Germany.

\clearpage
\newpage

\end{document}